\documentclass[12pt]{article}
             
\title{On the Lorentz Transformations of Momentum and Energy.}
\author{M. Toller \thanks{e-mail: toller@iol.it}\\ 
via Malfatti n. 8  \\
I-38100 Trento, Italy}

\begin{document} 
\maketitle
                 
\begin{abstract}
Motivated by ultra-high-energy cosmic ray physics, we discuss all the possible alternatives to the familiar Lorentz transformations of the momentum and the energy of a particle. Starting from natural physical requirements, we exclude all the possibilities, apart from the ones which arise from the usual four-vector transformations by means of a change of coordinates in the mass-shell. This result confirms the remark, given in a preceding paper, that, in a theory without preferred inertial frames, one can always define a linearly transforming energy parameter to which the GZK cutoff argument can be applied. We also discuss the connections between the conservation and the transformation properties of energy-momentum and the relation between energy-momentum and velocity. 

\end{abstract}

\section{Introduction.}

In the last few years several authors have suggested  that the relativity principle and the Lorentz symmetry might not be valid for processes that involve ultra-high-energy particles and that in this way one could explain some unexpected observations in cosmic ray physics \cite{ACEMNS,CK,Kifune,CG2,GM,BC,SG,ACP,KL}. 

It has also been suggested that these new phenomena can also be explained  maintaining the validity of the relativity principle and of the Lorentz symmetry, but assuming that the Lorentz group acts in a non-linear way on the energy and on the components of momentum \cite{AC,MS,KG}. 

This possibility, sometimes called ``doubly special relativity'', has subsequently been discussed in many articles. A list of references and a summary of some results are given in ref.\ \cite{AC2}. Many authors have remarked that, in the proposed models, the non-linearly transforming variables are non-linear functions of the usual linearly transforming variables. More recently \cite{LN,KGN,KGN2,Toller,AKD,GKV} it has been observed that a simple change of variables cannot affect the final results of a calculation concerning, for instance, the kinematics of particle collisions. As a consequence, at least in some important respects, the proposed ``doubly special relativistic'' theories are not physically distinguishable from the usual special relativistic formalism.

In order to discuss the possible existence of non-linearly transforming kinematic variables that are not functions of a linearly transforming four-vector, in ref.\ \cite{Toller} we have adopted an abstract point of view, namely we have considered these four quantities as a (redundant) set of coordinates on a three-dimensional manifold $\Pi$, usually called the mass-shell. In differential geometry, it is a familiar fact that the important properties of the geometric objects do not depend on the choice of the coordinates. We have assumed that the Lorentz group acts continuously and transitively on $\Pi$, namely that $\Pi$ is an homogeneous space. In the present paper we adopt this point of view and we derive some of the results of ref.\ \cite{Toller} starting from more general assumptions.

Note that we only consider the transformation properties of energy and momentum on the mass shell. This is sufficient for a simple phenomenological treatment of the kinematics of particle propagation and collisions. For more theoretical considerations it is necessary to investigate the non-transitive action of the Lorentz group on the whole four-dimensional energy-momentum space. This harder problem is discussed in ref. \cite{KGN2}.

In the usual relativistic theory, the mass-shell $\Pi$ of a massive particle is defined in terms of the components of a four-vector $\pi^i = (\pi^0, \vec\pi)$ by  the equation
\begin{equation} \label{Homo1}
\pi_i \pi^i = (\pi^0)^2 - |\vec\pi|^2 = 1, \qquad \pi^0 \geq 1.
\end{equation}
Note that the quantities $\pi^i$, which transform linearly under the Lorentz group, form a (redundant) set of coordinates on $\Pi$, but we do not identify them with the components of the four-momentum. In a similar way, for a massless particle the manifold $\Pi$ is defined by
\begin{equation} \label{Homo0}
\pi_i \pi^i = 0, \qquad \pi^0 > 0.
\end{equation}

It is a remarkable experimental fact that one can assign to every kind of particle a parameter $m$ in such a way that the four-momentum defined by
\begin{equation} \label{Mom}
p^i = m \pi^i
\end{equation}
is conserved in collisions, at least if the energies lie in the range up to now accessible in laboratory experiments. For massive particles $m$ is the mass and for massless particles we can always put $m = 1$, by means of a rescaling of the variables $\pi^i$ that satisfy the dilatation invariant constraint (\ref{Homo0}).

As we have said above, a ``trivial'' deformation of the Lorentz transformations of the four-momentum of a particle can be introduced by modifying the relations (\ref{Mom}), namely by introducing new coordinates in the same manifold $\Pi$. A non-trivial deformation replaces the abstract homogeneous space $\Pi$ by another (non isomorphic) one. The second possibility has been examined in ref.\ \cite{Toller} with a negative result. However, a particular definition of ``small'' deformation was adopted there and with a more general definition the result could have been different. Here we want to settle this argument by considering all the mathematically possible three-dimensional homogeneous spaces, up to isomorphisms.

\section{Three-dimensional homogeneous spaces.}

 A detailed classification of all the homogeneous spaces (with dimension smaller than six) of the Lorentz group is given in ref.\ \cite{Toller1} and here we summarize the results relevant for our problem. Some of the three-dimensional homogeneous spaces are transitive with respect to the action of the rotation subgroup. This means that any possible four-momentum of the particle can be obtained from another arbitrary four-momentum by means of a rotation. As a consequence, it is not possible to define a rotation-invariant energy. As it has been discussed in ref.\ \cite{Toller}, this is clearly a physically unacceptable feature and we disregard these spaces in the following. 

All the other three-dimensional homogeneous spaces are isomorphic to one of the following five spaces (one of them depending on a continuous parameter $k$).
\renewcommand{\labelenumi}{\alph{enumi})}
\begin{enumerate}
\item $\Pi_3^+$ defined by eq.\ (\ref{Homo1}).
\item $\Pi_3^0$ defined by eq.\ (\ref{Homo0}).
\item $\Pi_3^-$ defined by the equation
\begin{equation}  \label{Homo2}
\pi_i \pi^i = -1.
\end{equation} 
\item $\Pi_{3,+}^-$ obtained from the space $\Pi_3^-$ by identifying opposite points $\pm \pi^i$. One can also impose the condition $\pi^0 \geq 0$ and identify only  the points $(0, \pm \vec \pi)$.
\item $\Pi_{3,k}^0$ ($k > 1$) obtained from the space $\Pi_3^0$ by identifying points of the kind $k^m \pi^i$, where $m$ is an arbitrary integer.  One can also impose the condition $1 \leq \pi^0 \leq k$ and identify only the points $(1, \vec \pi)$ and $(k, k \vec \pi)$.
\end{enumerate}

The first three spaces are orbits in a four-vector space of the kind considerd by Wigner \cite {Wigner}. The quantities $\pi^i$ form a redundant set of global coordinates.  The last two spaces (which were not taken into account in ref.\ \cite{Toller}) cannot be considered as orbits in a four-vector space and the functions $\pi^i$, being multi-valued, cannot be considered as global coordinates. However, every point of $\Pi$ has a neighborhood in which a single value of these functions can be chosen in a continuous way and in this neighborhood they provide a redundant set of local coordinates, which transform as the components of a four-vector under Lorentz transformations sufficiently near to the identity.

It is important to remember that not all the points of the homogeneous spaces are (up to now) accessible to experimental investigation. It is clear that particles with extremely high energies are not available, but, for the massless particles, there are also problems with the extremely low energies corresponding to wave lengths larger than the planetary scale. We say that the points of $\Pi$ that have been subjected to a detailed experimental investigation form the ``accessible'' kinematic region.

It is clear that the first two homogeneous spaces describe rather well, in the accssible kinematic region, the known properties of the massive and massless particles, respectively. We want to show that the other three spaces are not relevant for physics. Presumably, they would introduce serious problems in the construction of a consistent physical theory, for instance a quantum field theory. However, we shall not use this argument, because the effects we are considering could be consequences of quantum gravity, a not yet well established theory. We prefer to base our discussion on simple phenomenological reasonings, directly related to the experimental observations.

\section{Conservation and transformation laws.}

Our arguments are mainly based on the conservation of energy and momentum. As we have already said, the experiments suggest that that in a given open connected region of $\Pi$ (the accessible kinematic region) one can define four functions $p^i$, interpreted as the energy and the components of the momentum, which are conserved. This means that if we consider the process $A + B \to C + D$, the equalities 
\begin{equation} \label{Cons}
p^i(\pi_A) + p^i(\pi_B) = p^i(\pi_C) + p^i(\pi_D)
\end{equation}
are satisfied if $\pi_A, \pi_B, \pi_C$ and $\pi_D$ belong to the accessible kinematic region. We also assume that the energy $p^0$ is a scalar, the momentum $\vec p$ is a vector under rotations and the accessible kinematic region is rotationally invariant. Then we can put
\begin{equation} \label{EnMom}
p^0 = f(\pi^0), \qquad  \vec p = g(\pi^0) \vec\pi, \qquad a < \pi^0 < b.
\end{equation}

As we have reminded in ref.\  \cite{Toller}, it has been proven a long time ago \cite{LT} that the conservation of energy and momentum together with their transformation properties under rotations and other minor additional assumptions imply eq.\  (\ref{Mom}). This result is in agreement with the assumptions adopted in ref.\ \cite{JV}, namely that the linearly transforming quantities (\ref{Mom}) are conserved in collisions, while the non-linarly transforming quantities (which are considered physically more significant) satisfy more complicated constraints.

However, the proofs that can be found in the textbooks \cite{Moller,Jackson} are not sufficiently general for our purposes, since they are given only for massive particles and assume that both the coordinates $\pi^i$ and $p^i$ are defined on the whole homogeneous space $\Pi$. Here we give a simple more general proof, which holds for all the five homogeneous spaces introduced above and assumes the conservation laws only in the interval $a < \pi^0 < b$. 

It is sufficient to assume that energy and momentum are conserved in the elastic scattering of identical particles. One can also consider only ``peripheral''  scattering, namely final states that differ only slightly from the initial state. Besides eqs.\ (\ref{Cons}) and (\ref{EnMom}), we only assume that the function $f$ has a continuous second derivative and that the function $g$ has a continuous first derivative.

If we consider the particular initial and final states defined by
\begin{displaymath}
\pi_A = (\omega, 0, 0, \eta), \qquad
\pi_B = (\omega, 0, 0, -\eta),
\end{displaymath}
\begin{equation}
\pi_C = (\omega, 0, \eta \sin\theta, \eta \cos\theta), \qquad
\pi_D = (\omega, 0, -\eta \sin\theta, -\eta\cos\theta),
\end{equation}
it follows from eq.\ (\ref{EnMom}) that the conservation equation (\ref{Cons}) is satisfied.

If we apply to all the variables a boost along the third axis with a small rapidity $\zeta$, in such a way that it does not take the dynamical variables outside the open region where they are defined, we obtain
\begin{displaymath}
\pi'_A = (\omega \cosh\zeta + \eta \sinh\zeta, 0, 0, \omega \sinh\zeta +\eta \cosh\zeta), \qquad
\end{displaymath}
\begin{displaymath}
\pi'_B = (\omega \cosh\zeta - \eta \sinh\zeta, 0, 0, \omega \sinh\zeta - \eta \cosh\zeta), \qquad
\end{displaymath}
\begin{displaymath}
\pi'_C = (\omega \cosh\zeta + \eta \cos\theta \sinh\zeta, 0, \eta \sin\theta, 
\omega \sinh\zeta +\eta \cos\theta \cosh\zeta), \qquad
\end{displaymath}
\begin{equation}
\pi'_D = (\omega \cosh\zeta - \eta \cos\theta \sinh\zeta, 0, -\eta \sin\theta, 
\omega \sinh\zeta - \eta \cos\theta \cosh\zeta).
\end{equation}

The conservation equations for the energy and the third component of momentum take the form
\begin{displaymath}
f(\omega \cosh\zeta + \eta \sinh\zeta) + f(\omega \cosh\zeta - \eta \sinh\zeta) =
\end{displaymath}
\begin{equation} \label{En}
= f(\omega \cosh\zeta + \eta \cos\theta \sinh\zeta) + f(\omega \cosh\zeta - \eta \cos\theta \sinh\zeta),
\end{equation}
\begin{displaymath}
(\omega \sinh\zeta +\eta \cosh\zeta) g(\omega \cosh\zeta + \eta \sinh\zeta) +
\end{displaymath}
\begin{displaymath}
+ (\omega \sinh\zeta - \eta \cosh\zeta) g(\omega \cosh\zeta - \eta \sinh\zeta) =
\end{displaymath}
\begin{displaymath}
= (\omega \sinh\zeta +\eta \cos\theta \cosh\zeta) g(\omega \cosh\zeta + \eta \cos\theta \sinh\zeta) +
\end{displaymath}
\begin{equation} \label{Momen}
+ (\omega \sinh\zeta - \eta \cos\theta \cosh\zeta) g(\omega \cosh\zeta - \eta \cos\theta \sinh\zeta).
\end{equation}

We calculate the second derivatives of both sides of eq.\ (\ref{En}) and the first derivatives of both sides of eq.\ (\ref{Momen}) and we put $\zeta = 0$. If $\eta > 0$ and $\theta$ is small, but not vanishing, we obtain the result
\begin{equation} 
f''(\omega) = 0, \qquad  g'(\omega) = 0.
\end{equation}
We note that $\eta = |\vec\pi|$ can vanish only for $\omega = \pi^0 = 1$, if $\Pi = \Pi_3^+$.  It follows that $\vec p = m \vec\pi$ and $p^0 = a \pi^0 + b$. 

The quantities $m, a$ and $b$ depend on the kind of particle and can be determined, up to ambiguities due to the choice of the units, by considering low energy processes that involve many kinds of particles.  From the conservation of $p^0$ and the Lorentz transformation law of $\pi^i$, it follows that $a \vec\pi$ and $b$ are also conserved quantities. In order to avoid the conservation of two different vector quantities, we have to put (with a convenient choice of the units) $a = m$. The scalar conserved quantity $b$ (for instance, proportional to the electric charge), can be eliminated by changing the arbitrary additive constant of the energy, and the proof of eq.\ (\ref{Mom}) is complete.

From eq.\ (\ref{Mom}) and eqs.\ (\ref{Homo1}), (\ref{Homo0}), (\ref{Homo2}), we obtain dispersion laws of the kind
\begin{equation} 
p^i p_i = C,
\end{equation} 
where $C = m^2$ in the case a), $C = -m^2$ in the cases c) and d) and $C = 0$ in the cases b) and e). We have shown in a very general way that, as a consequence of the relativity principle, in the kinematic region in which the energy-momentum conservation laws are valid, the dispersion law must have the usual quadratic Lorentz invariant form. We have also seen that only the usual mass-shell $\Pi_3^+$ can be used to describe massive particles.  

\section{Massless particles.}

For massless particles, the situation is more delicate. From the experiments one can only deduce upper bounds to the absolute value of the constant $C$ which appears in the dispersion law and all the five homogeneous spaces are admitted from this point of view. A discussion of the upper bounds to the photon mass can be found in ref.\ \cite{Jackson}. 

In the case c), negative values of $\pi^0$ are allowed, and, in order to avoid negative values of $p^0$, which are not observed, one has to modify eq.\ (\ref{Mom}). For instance, one can use the formula
\begin{equation} 
p^0 =  \frac m 2 \left(\pi^0 + ((\pi^0)^2 + 1)^{1/2}\right), \qquad
\vec p = p^0 |\vec\pi|^{-1} \vec\pi,
\end{equation} 
which coincides with  eq.\ (\ref{Mom}) for $p^0 \gg m$. For small values of $p^0$, the quantities $p^i$ are not given by eq.\  (\ref{Mom}) and cannot be conserved. Note that both negative values of the energy and energy non-conservation are hardly compatible with a thermal equilibrium state of the radiation.

In the case d), if $g(0) > 0$, the momentum $\vec p$ is discontinuous for $\pi^0 = 0$  and cannot be conserved in the very low energy region. In fact, one can imagine situations in which one of the terms in eq.\ (\ref{Cons}) is discontinuous, while the other three are continuous, which is a contradiction. If $g(0) = 0$, $g(\pi^0)$ cannot be constant and the conservation laws are again violated. A similar situation appears in the case e), in which  the quantities $p^i$ are discontinuous at $\pi^0 = m$ (or at $\pi^0 = k m$), unless we have $f(m) = f(km)$ and $g(m) = k g(km)$.

In conclusion, we cannot exclude that the homogeneous spaces $\Pi_3^+$, $\Pi_3^-$, $\Pi_{3+}^-$ and $\Pi_{3,k}^0$ can be used to decribe, for instance, the photon, but $m$ has to lie well below the accessible energy region. In the case $\Pi = \Pi_{3,k}^0$, moreover, we have to require that $k m$ lies well above the accessible energy region. If these conditions are satisfied, these homogeneous spaces are experimentally indistinguishable form the space $\Pi_3^0$, which is traditionally adopted to describe massless particles.

The anomalies which could appear at energies of the order of the parameter $m$ concern the behavior of the
long-wave electromagnetic waves, namely some modifications of the classical Maxwell equations. Of course, it is difficult to test the Maxwell equations at a distance scale larger than the planetary scale \cite{Jackson}.

In cosmic ray physics, one is more interested in ultra-high-energy phenomena. In this energy range, the homogeneous spaces $\Pi_3^+$, $\Pi_3^-$ and $\Pi_{3+}^-$ are practically indistinguishable from the space $\Pi_3^0$ and they represent just a useless complication in the description of high energy massless particles. 

The case $\Pi = \Pi_{3,k}^0$ deserves more attention. If $f(\pi^0)$ is continuous, it cannot be a monotonic function and there is more than one state with the same energy $p^0$ and the same direction of $\vec p$. One has to explain why these additional states are not observed in laboratory experiments. If $f(\pi^0)$ is monotonic, we must have $f(km) \gg f(m)$ and, if in a collision a particle with energy slightly larger than $f(km)$ is produced, it appears as a particle with energy near to $f(m)$, namely, in practice, it becomes unobservable. In a similar way, if in a low energy process a particle with energy smaller than $f(m)$ is emitted, it appears as a particle with energy near to $f(km)$, namely as a ultra-high-energy particle. 

These predictions look rather incredible, even if the probability of these events can be made as small as we like by choosing the parameters $m$ and $k$.  In any case, these strange features do not help in any way the explanation of the cosmic ray anomalies which have motivated our investigation.

All these considerations suggest that the only abstract homogeneous spaces which can reasonably be used for the description of particles are the hyperboloid $\Pi_3^+$ and the light-cone $\Pi_3^0$, in agreement with the conclusions of ref.\ \cite{Toller}. 

\section{Covariance and average velocity.}

Another important property of the particles, besides their four-momentum, is their velocity. We do not mean the istantaneous velocity, which could be badly defined if the space-time has some kind of ``granular'' structure, but the average velocity over a time interval $t$ much larger than the scale $t_0$ of this structure, probably the Plank scale. If the time interval $t$ is also much larger than $\hbar / |\vec p|$, the average velocity can be treated as a classical (non quantum) observable. It is operationally well defined if its value does not vary appreciably in the time interval $t$, namely if the external fields, including the curvature, are not too large and we exclude the collisions with other particles.

Under these conditions, we can describe the velocity by means of a ray in the Minkowski space-time (a stright line passing through the origin) parallel to the tangent to the world-line of the particle. In other words, the velocity space is the three-dimensional projective space corresponding to the Minkowski spacetime. With respect to the action of the Lorentz group it splits into three orbits. The time-like rays cross the hyperboloid (\ref{Homo1}) in one point and form an homogeneous space isomorphic to $\Pi_3^+$. The space-like rays cross the hyperboloid (\ref{Homo2}) in two opposite points and form an homogeneous space isomorphic to $\Pi_{3,+}^-$. The light-like rays form a two-dimensional orbit diffeomorphic to a sphere, called the {\it celestial sphere}. In ref.\ \cite{Toller1} it is indicated by $\Pi_2$ and it is the only two-dimensional homogeneous space.

If the relativity principle is valid, there is a covariant mapping between the homogeneous space that describes the energy-momentum of a particle and the velocity space. The image of this mapping is one of the three orbits mentioned above. It has been shown in ref.\ \cite{Toller1} that this mapping is uniquely determined by the covariance. $\Pi_3^+$ is trivially mapped onto itself, $\Pi_3^-$ is mapped onto $\Pi_{3,+}^-$, $\Pi_{3,+}^-$ is mapped onto itself, $\Pi_3^0$ and $\Pi_{3,k}^0$ are mapped onto the celestial sphere  $\Pi_2$. 

This means that, as a consequence of the relativity principle, the average velocity depends on the kinematic variables $\pi^i$ in the usual way and this dependence is not affected by a possible non-linear choice of the energy-momentum coordinates $p^i$. This conclusion was also reached in refs.\ \cite{KM,KR}. If we require that the velocity is given by the familiar Hamilton equation, the coordinates $p^i$ cannot be chosen arbitrarily. It also follows that it is difficult to explain a possible dependence on the energy of the average velocity in vacuum of gamma rays without abandoning the relativity principle, namely without introducing privileged inertial frames (see also \cite{LN2,THMT}).

\section{Conclusions.}

An important consequence of our analysis is that, as a consequence of the relativity principle, in the kinematic calculations one can always use the components of energy-momentum defined by eq.\ (\ref{Mom}), which satisfy the usual dispersion law and are conserved in collisions with a center-of-mass energy in the range studied in laboratory experiments. Of course, one can use different coordinates, but the final result cannot change, since the effect of a modification of the dispersion law is compensated by a modification of the conservation laws.

This remark can be applied to the treatment of the GZK cutoff \cite{ZK,Greisen}, which arises because protons with an energy larger than $5 \times 10^{10} \, GeV$ colliding with the photons of the cosmic microwave background (CMB) reach the threshold for pion photoproduction and cannot cover a long intergalactic distance without loosing most of their energy.

The collisions between ultra-high energy protons and the CMB photons have a low center-of-mass energy and, as we have seen, can be treated by means of the usual kinematic formalism. It is also correct to use the cross-sections measured in laboratory experiments and the argument leading to the GZK cutoff for the energy coordinate defined by eq.\ (\ref{Mom}) cannot be avoided.

In conclusion, the present more general analysis confirms  the results of ref.\ \cite{Toller}, namely that if the arrival of cosmic rays with energy beyond the GZK cutoff \cite{Takeda} is confirmed (see, however, ref.\ \cite{BW} and references therein), we have to consider the following alternatives.
\renewcommand{\labelenumi}{\Alph{enumi})}
\begin{enumerate}
\item The cosmic rays do not come from the far intergalactic space.
\item The relativity principle is valid and the energy defined by eq.\ (\ref{Mom}) is subject to the GZK cutoff, but it is not the energy evaluated in the experiments. This may happen because the collisions of the cosmic rays with the nuclei in the high atmosphere, which have a center-of-mass energy much larger than the collisions studied in the terrestrial laboratories, have unexpected features, possibly a breakdown of the conservation laws.
\item  The relativity principle is not valid and there are privileged inertial frames. The fundamental laws may still be Lorentz covariant, but some long-range vector or tensor field may have a non vanishing expectation value that singles out the privileged frames.
\end{enumerate}

\end{document}